# Noise Suppression with Cryogenic Resonators

Eugene N. Ivanov and Michael E. Tobar

*Abstract -* **We show that fast technical fluctuations of microwave signals can be strongly suppressed by cryogenic resonators. The experiments were carried out with sapphire resonators cooled to approximately 6 K at frequencies around 11 GHz. Each sapphire crystal was shaped like a spindle with the rotational and crystal axes aligned within a degree. Noise suppression factors in excess of 50 dB were measured at 10 kHz offset from the carrier. This was achieved by rotating the sapphire spindle and moving the coupling probes relative to its surface. Microwave signals with reduced fast technical fluctuations allow the high-precision tests of fundamental physics, as well as the development of better radars.**

*Index Terms–* **cryogenic resonators, noise suppression**

## I. INTRODUCTION

THE progress in laser frequency stabilization and optical frequency synthesis resulted in the generation of extremely frequency-stable microwave signals [1-6]. Recently, NIST scientists have synthesized microwave signals with the fractional frequency instability of $1 \times 10^{-18}$ at $10^4$s of averaging from the Ytterbium cold atom clock [7]. Microwave signals with very low phase noise were also synthesized from the optical sources [8-11]. For example, reference [11] describes the optical synthesis of the 12 GHz signals with Single-Sideband (SSB) phase noise spectral density $\mathcal{L}_\phi \approx -173\ dBc/Hz$ at Fourier frequency $F$=10 kHz. Such a noise performance is comparable to classical oscillators based on the Sapphire Loaded Cavity (SLC) resonators operating at room temperatures [12].

The SLC resonators offer the highest Q-factors at microwave frequencies. When cooled to 4K, their intrinsic Q-factors may exceed a few billion [13], enabling efficient filtering of the transmitted signals' fast technical fluctuations. This work describes how to take advantage of the high Q-factors of the cryogenic SLC resonators to generate spectrally-pure microwaves.

To characterize the noise suppression efficiency of the resonator, we introduce its Noise Suppression Factor (NSF) as a ratio of phase noise power spectral densities of the transmitted and incident signals: $NSF = S_\varphi^{trans}/S_\varphi^{inc}$. For a resonator with Lorentzian transfer function and resonant frequency tuned to that of the incident signal, an analytical expression for the NSF can be derived as

$$NSF\,(F) = 1 \Big/ \big(1 + (F/\Delta f_{0.5})^2 \big)$$

where $2\Delta f_{0.5}$ is the resonator 3 dB linewidth.

For a typical cryogenic SLC resonator with the linewidth of 50 Hz, we obtain NSF ≈ 52 dB at $F$ = 10 kHz. This is sufficient to reduce technical fluctuations of any commercial quartz-based microwave synthesizer to the standard thermal noise limit [14], provided that the power of the transmitted signal extracted from the cryostat is close to 1 mW (see Fig. 7 below).

Transfer functions of real resonators are rarely Lorentzian. For example, the cryogenic SLC resonators' high-Q modes often exist as doublets [13, 15]. In such a case, the NSF needs to be either measured or computed. Below we discuss some results of our experiments with two cryogenic SLC resonators, including the measurements of their transmission coefficients and noise suppression factors.

## II. SAPPHIRE RESONATOR

Fig. 1 shows the SLC resonator's key parts: the sapphire spindle, the copper shield, and two lids. The sapphire spindle is held at the center of the bottom lid with a copper collet. Annular slots in the bottom lid enable it to be rotated relative to the coupling probes (not shown in Fig. 1). The top lid is used for attaching the resonator to the cold finger of the cryocooler.

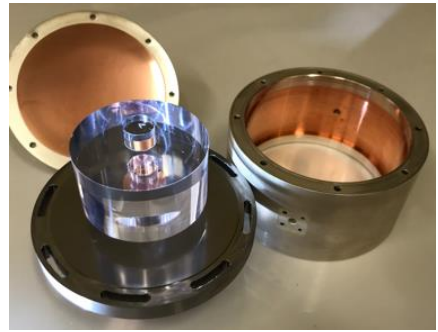

Fig. 1. Sapphire Loaded Cavity (SLC) resonator

Fig. 2 shows the amplitude transmission coefficients of the 11.201 GHz sapphire resonator measured at four different angular positions of the sapphire spindle. All measurements were carried out at 6.5 K.

The shape of the resonant peak in Fig. 2c is almost Lorentzian. A slight asymmetry is due to the spurious low-Q mode's proximity whose electromagnetic field is poorly confined to the crystal. For that reason, moving the coupling probes away from the crystal surface increased the resonance peak asymmetry and decreased its contrast.

¹ Eugene Ivanov, School of Physics,
ARC Centre of Excellence for Engineered Quantum Systems, The University of Western Australia, 35 Stirling Highway, Crawley, 6009
(eugene.ivanov@uwa.edu.au)

Michael Tobar, School of Physics, ARC Centre of Excellence for Engineered Quantum Systems, The University of Western Australia, 35 Stirling Highway, Crawley, 6009
(michael.tobar@uwa.edu.au)

Our study of the cryogenic SLC resonators shows that:
1. It is possible to eliminate one of the components of the doublet resonance by crystal rotation;
2. The lineshape of the singlet resonance can closely match Lorentzian with the contrast in excess of 50 dB;
3. The *intrinsic* Q-factors of the individual resonances are invariant to crystal rotations (the *loaded* Q-factors of the resonances in Fig. 2 varied from 2 billion to 200 million, whereas their *intrinsic* Qs remained close to 2 billion).

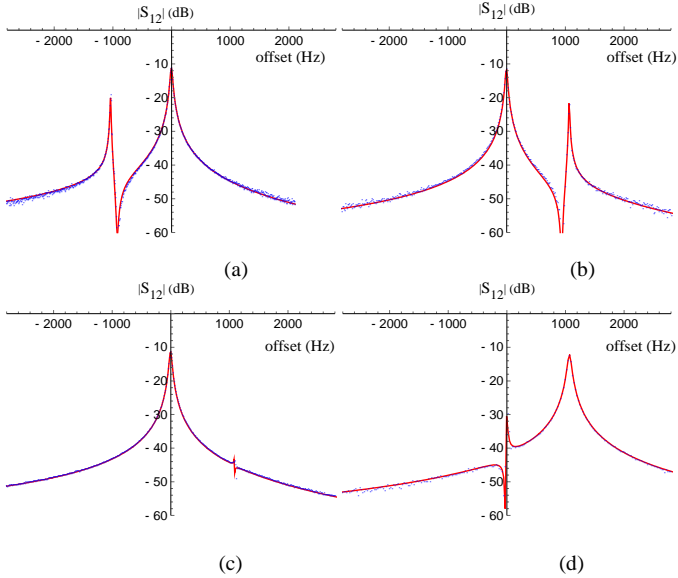

Fig. 2. Amplitude transmission coefficients of the 11.201 GHz cryogenic SLC resonator at different angular positions of sapphire spindle:
(a) Φ = - 4°, (b) Φ = - 2°, (c) Φ = - 1°, (d) Φ = 4°.

## III. NOISE SUPPRESSION MEASUREMENTS

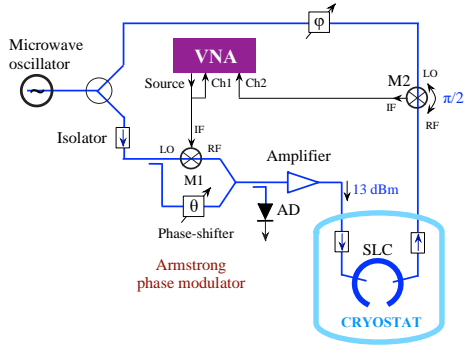

Fig. 3. Noise suppression measurement system: AD – amplitude detector, VNA-vector network analyzer, M1, M2 – double-balanced mixers

Fig. 3 shows a schematic diagram of the experimental setup for measuring the cryogenic resonators' noise suppression factors. It features two key components: (i) mixer-based phase-sensitive readout system and (ii) Armstrong phase modulator [16]. The latter resembles the Mach-Zehnder interferometer, except that it contains a double-balanced mixer (M1) in one of its arms. An amplitude detector at the modulator's output (AD in Fig. 3) acts as a sensor of spurious AM-modulation. If, for example, a sinewave at frequency Ω drives the modulator, and no response at the Ω is observed at the detector's output, then a pure PM-modulated signal is generated. The reasons for choosing Armstrong modulator instead of the conventional voltage-controlled phase-shifter were: (i) much broader modulation bandwidth and (ii) reduced intensity of spurious amplitude modulation [17, 18].

To measure the NSF over a broad range of Fourier frequencies, we drive the Armstrong modulator with a "white" voltage noise from the Vector Network Analyzer (VNA in Fig. 3). This results in a microwave signal with a random phase modulation incident on the cryogenic resonator. Next, we tune our readout system to be phase-sensitive. For that, we set two microwave signals entering mixer M2 in quadrature by varying phase delay φ of the reference signal. In such a regime, the mean voltage at the mixer M2 output is close to zero, and voltage fluctuations are synchronous with phase fluctuations of the transmitted microwave signal.

The VNA in Fig.3 measures the ratio of two spectral densities: $R(F) = S_{u1,2}(F)/S_{source}$, where $S_{u1,2}$ is the cross-spectral density between voltage fluctuations applied to the Armstrong modulator and extracted from the phase detector, and $S_{source}$ is the spectral density of voltage noise applied to the Armstrong modulator. For the cross-spectral density $S_{u1,2}$, it can be shown that

$$S_{u1,2}(F) = S_{source} K_{PD} \frac{d\phi}{du} \rho(F)$$

where $K_{PD}$ is the phase detector sensitivity, and $d\phi/du$ is the voltage-to-phase conversion efficiency of the modulator. The last factor in the above equation is a complex function of Fourier frequency whose modulus squared is the noise suppression factor: $|\rho(F)|^2 = NSF(F)$. This means that the resonator noise suppression factor is a normalized frequency response measured by the VNA, i.e., $NSF(F) = |R(F)/R(0)|^2$.

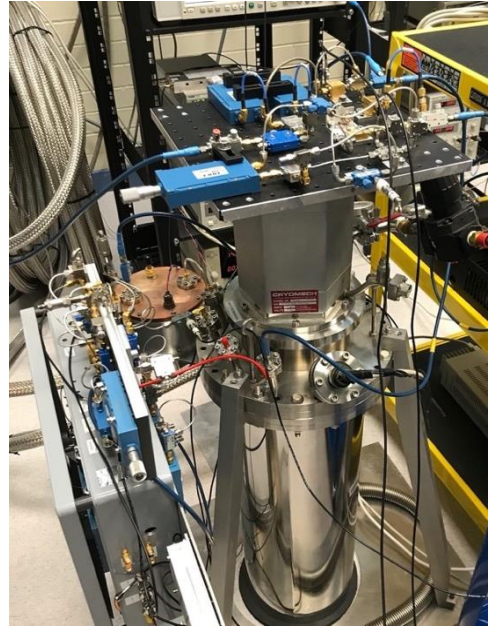

Fig. 4. Experimental setup used for measuring noise suppression factor of the cryogenic resonator

Fig. 4 shows the experimental setup for measuring the noise suppression factor of the cryogenic SLC resonator. The resonator is kept at 6.5 K inside the stainless-steel cryostat of the pulse tube refrigerator. Microwave components of the noise measurement system are assembled on the optical breadboard on top of the pulse tube cold head. Various microwave components to the cryostat left are part of the low-noise oscillator [19]; they are not involved in the NSF measurements.

Fig. 5 shows the NSF as a function of Fourier frequency for the resonator with the transmission coefficient in Fig. 2c. A solid line in Fig. 5 is the computed result. A small discrepancy between the calculated and measured data at F > 30 kHz is due to the numerical model's limitations to calculate the NSF [20].

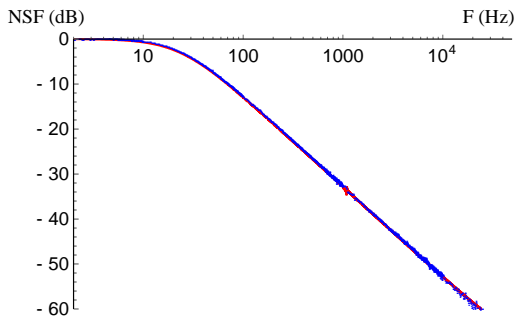

Fig. 5.  Noise suppression factor of 11.201 GHz SLC resonator

As mentioned earlier, two cryogenic SLC resonators were characterized both in terms of their transmission coefficients and noise suppression factors. This characterization for the second resonator operating at frequency 11.343 GHz is presented in Fig. 6.

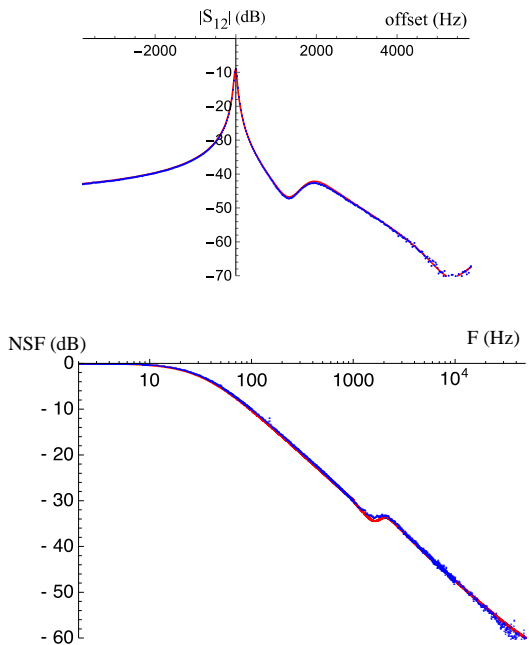

Fig. 6.  Transmission coefficient (top) and noise suppression factor (bottom) of the SLC resonator operating at 11.343 GHz.

A step-like distortion at the NSF's dependence on Fourier frequency around 2 kHz is due to the resonator's spurious low-Q mode. At Fourier frequencies below 1 kHz, the noise suppression factor measured is identical to that of the Lorentzian resonator with the linewidth of 58 Hz.

Fig. 7 illustrates the efficiency of noise suppression by the cryogenic sapphire resonator. The top trace in Fig. 7 is the SSB phase noise spectrum of the microwave frequency synthesizer (Keysight E8257D). The phase noise was measured at 11.2 GHz using the two-oscillator technique described in [21].

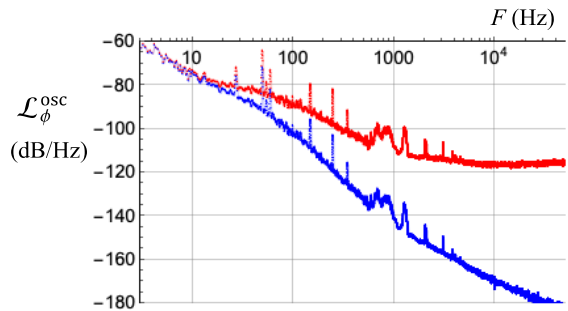

Fig. 7.  SSB phase noise spectra of the E8257D at 11.2 GHz: top trace – measured phase noise of the incident signal; bottom trace is the inferred phase noise of the transmitted signal.

The bottom trace in Fig. 7 is the inferred phase noise of a signal transmitted through the resonator with the amplitude transfer function shown in Fig. 2c. At Fourier frequencies above 30 kHz, the SSB phase noise of the transmitted signal approaches -180 dBc/Hz.  This is close to the limit set by thermal fluctuations for the transmission signal with a power of 1 mW. The direct measurements of such extremely weak phase fluctuations are a subject of on-going research.

## IV. CONCLUSION

Our results demonstrate that cryogenic sapphire resonators enable a significant improvement in spectral purity of generated microwave signals. Furthermore, it's worth emphasizing that both phase and amplitude fluctuations of the transmitted signal are reduced by the same amount. This is because the resonator in transmission acts as a band-pass filter that suppresses any fluctuations outside its bandwidth regardless of their nature.

The low-pass filtering of amplitude fluctuations by the cryogenic resonators allows more accurate measurements of weak phase fluctuations. The problem is that until recently, little attention has been paid to the cancellation of oscillator amplitude fluctuations. With the rapid progress in oscillator frequency stabilization, the situation changed, and amplitude noise is often the key factor limiting phase noise measurements' accuracy.

Authors acknowledge useful discussions with NIST, Boulder Laboratories researchers J. Bergquist, D. Hume, C. Oates, S. Diddams and F. Quinlan, as well as financial support of the Australian Research Council (grants DP190100071 and CE170100009)